\documentclass[twocolumn,preprintnumbers,showpacs,pre]{revtex4}
\usepackage{graphicx}
\usepackage{dcolumn}
\usepackage{bm}

\def\be{\begin{equation}}
\def\ee{\end{equation}}
\def\bea{\begin{eqnarray}}
\def\eea{\end{eqnarray}}
\def\ba{\begin{array}}
\def\ea{\end{array}}

\def\0{$\sigma_0$}

\def\t{\theta}

\begin{document}

\title{Flat lens without optical axis: Theory of imaging}
\author{W. T. Lu}
\email{w.lu@neu.edu}
\author{S. Sridhar}
\email{s.sridhar@neu.edu}
\address{Department of Physics and Electronic Materials Research Institute,
Northeastern University, Boston, MA 02115}

\date{\today }

\begin{abstract}
We derive a general theory for imaging by a flat lens without optical axis.
We show that the condition for imaging requires a material having \emph{%
elliptic dispersion relations with negative group refraction}. This medium
is characterized by two intrinsic parameters $\sigma $ and $\kappa $.
Imaging can be achieved with both negative and positive wave vector
refraction if $\sigma $ is a positive constant. The Veselago-Pendry lens is
a special case with $\sigma =1$ and $\kappa =0$. A general law of refraction
for anisotropic media is revealed. Realizations of the imaging conditions
using anisotropic media and inhomogeneous media, particularly photonic
crystals, are discussed. Numerical examples of imaging and requirements for
sub-wavelength imaging are also presented.
\end{abstract}


\maketitle

Since antiquity, the positive index of refraction of conventional materials
has required the use of curved surfaces to focus light. However there has
been a continuing quest for lenses with flat surfaces as they confer a
variety of advantages. Notable examples are the Fresnel lens and Maxwell's
fish eye lens \cite{Maxwell,BornWolf}. The Fresnel flat lens uses a
gradient-index $n(x)$ material (GRIN) \cite{GRIN,Wilkinson,Smith05} and hence must
possess an optical axis, i.e. is not translationally invariant along the
surface ($x$-axis).

The concept of negative refraction has led to new fundamental approaches as
well as applications in optics \cite{Tretyakov}. In 1968 Veselago \cite%
{Vesalago} pointed out that for a material with refractive index $n=-1$\ a
flat surface would focus light. A decade later, Silin \cite{Silin} discussed
a more general case and obtained the lens equation of a flat slab with
negative elliptic dispersion. Pendry's recent analysis \cite{Pendry}
demonstrating the possibility of sub-wavelength resolution with a flat slab
of such materials, as well as the experimental realization of the so-called
left-handed materials using composite media \cite{Shelby} and photonic
crystals (PhCs) \cite{Cubukcu,Parimi04} led to renewed interest in the
unique electromagnetic properties of these artificial materials. The unique
property of these flat lenses is the lack of optical axis. This type of flat
lenses can be realized in a PhC using negative refraction \cite{Luo02}
and has been demonstrated in microwave experiment \cite{Parimi03}. 
Super-resolution imaging through single negative index media are also studied
\cite{Platzman,Lu,Fang,Melville}. There is however
no complete theory of imaging by a flat lens which properly describes the
various features observed in the experiments and in numerical simulations.

In this paper, we present a general theory of imaging by a flat lens without
optical axis, resulting in a proposal for a new material which is
translationally invariant along the surface and has anisotropic refractive
index $n(\theta )$. Defining an optical phase condition for imaging, we show
that the condition requires a material having \emph{Elliptic dispersion
relations with Negative Group Refraction }(ENGRM) at the operating
frequency. The ENGRM is defined by two parameters:\ the anisotropy parameter 
$\sigma $ which is a measure of the ellipse eccentricity, and a phase factor 
$\kappa $ which determines the center of the ellipse. Refraction laws for
the wave vector and the group velocity for anisotropic media are derived.
The theory shows that ``perfect''\ images can be obtained, and \emph{imaging
is possible with negative as well positive refractive indices}. The required
conditions and consequences of flat perfect and imperfect lenses made of
ENGRMs are derived from a generalized Fermat's principle. Realization of flat
lenses using PhCs is discussed. In real PhCs, $\sigma $ is itself
angle-dependent in most cases, leading to some limitations for image
formation. Many of the proposals for PhC lenses are contained in the present
theory. The theory is also applied to recent experiments using PhCs and is
shown to successfully describe some intriguing aspects of the data.
Numerical simulations are presented that describe visual details of the
image formation in PhCs.

\begin{figure}[tbp]
\center{\includegraphics[angle=0,width=3.1in]{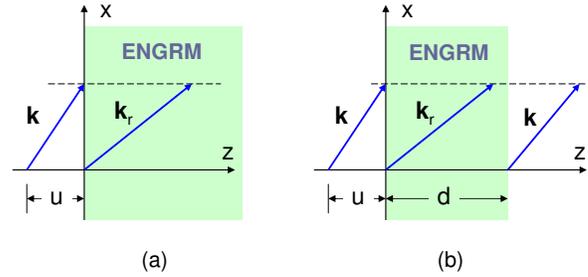}}
\caption{Wave vector refraction for (a)\ a single interface 
between the vacuum and an ENGRM and 
(b) an ENGRM slab.}
\label{fig1}
\end{figure}

\begin{figure}[tbp]
\center{\ \includegraphics [angle=0,width=3.1in]{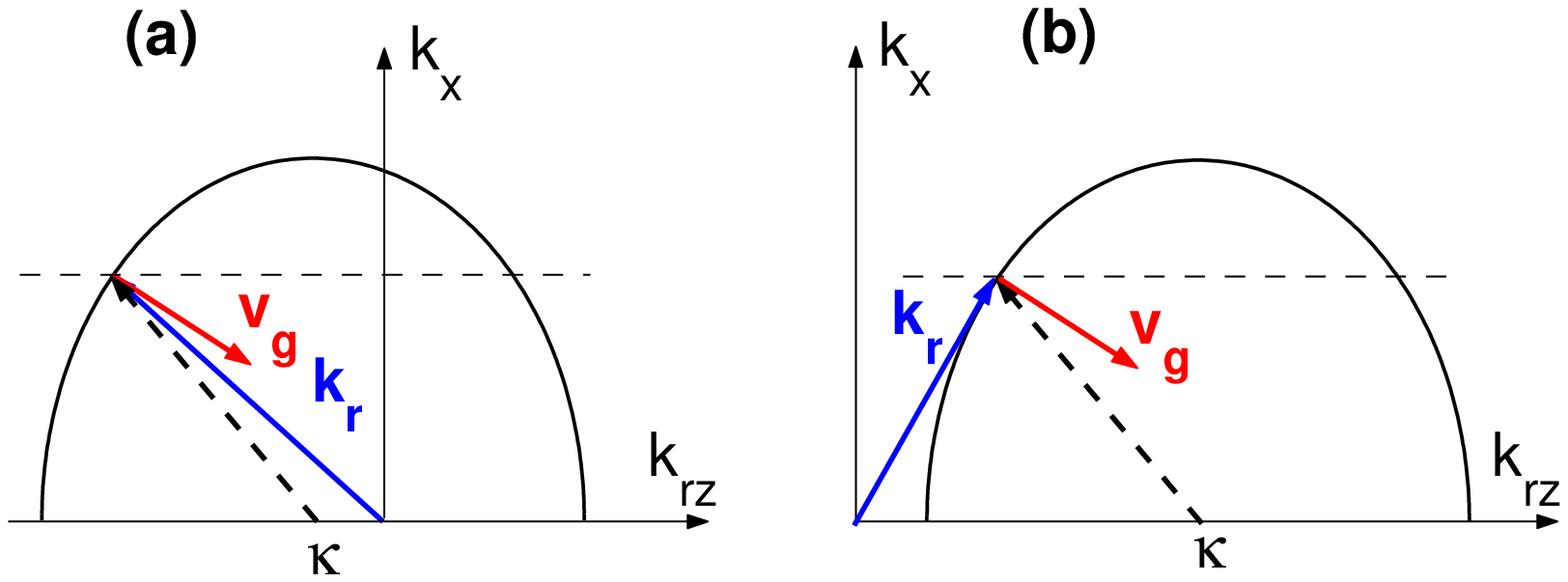}
\includegraphics [angle=0,width=3.0in]{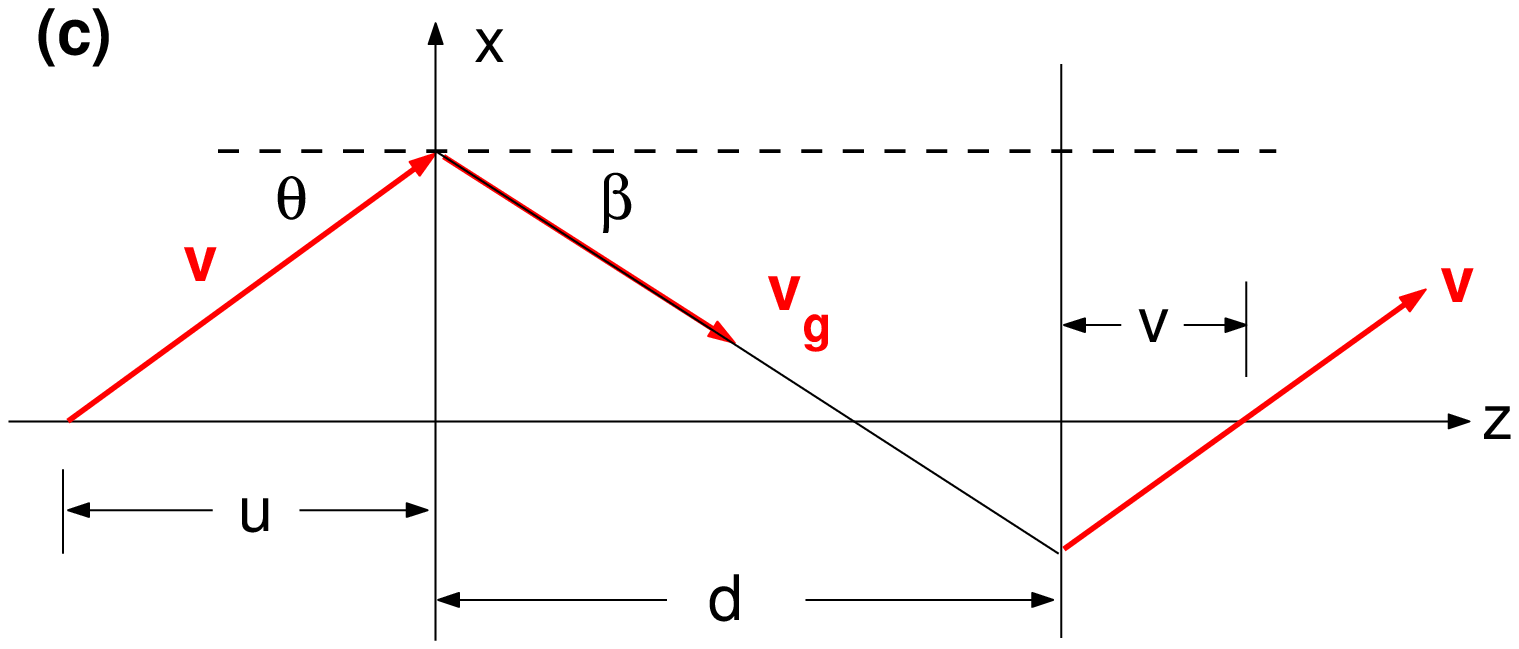}}
\caption{Group velocity for an ENGRM with elliptic EFS and $(\bf{k}_{r}-%
\protect\kappa \hat{z})\cdot \protect\nabla _{\bf{k}_{r}}\protect\omega %
<0$ and (a)\ negative or (b) positive wave vector refraction. The dashed
arrow is $\bf{k}_{r}-\protect\kappa \hat{z}$. (c) Ray diagram of the
group velocity for imaging by an ENGRM flat lens with 
$u+v=\protect\sigma d$.}
\label{fig2}
\end{figure}

We first consider the case of a single interface which is along the $x$-axis
and at $z=0$ as shown in Fig. \ref{fig1}(a). The wave vector $\mathbf{k}$ of 
an incident plane wave from
vacuum towards the interface makes an angle $\theta $ with the $z$-axis.
Here we consider the case that for any incident $\bf{k}$, there is only
one refractive wave vector ${\bf k}_{r}$. Continuity along the interface
requires that $k_{rx}=k_{x}$. A point source is at $x=0$ and $z=-u$ with $%
u>0 $. We now consider another point in the ENGRM at $x=\Delta x$ and $z=v$ with 
$v>0 $. The phase difference between these two points for each incident $%
\bf{k}$ is $\Phi =\Phi _{z}+\Phi _{x}$with $\Phi _{z}=k_{z}u+$ $k_{rz}v$
and $\Phi _{x}=k_{x}\Delta x$. According to Fermat's principle, the
formation of an image would require that the total phase $\Phi $ be
stationary. This restriction can be relaxed for a flat interface. Since the
surface is flat, any point of a finite-size source will be imaged to a point
with the same $x$-coordinate. Thus $\Delta x=0$ and $\Phi _{x}=0$. One only
needs to consider $\Phi _{z}$. A generalized Fermat's principle states that 
\emph{an image will be formed\ if the phase }$\Phi _{z}$\emph{\ is
stationary, }$\mathrm{d}\Phi _{z}/\mathrm{d}k_{z}=0$. The lens equation for
a single interface is 
\begin{equation}
u=\sigma v  \label{lens-0}
\end{equation}
with a material constant 
\begin{equation}
\sigma =-\mathrm{d}k_{rz}/\mathrm{d}k_{z}.  \label{sig-def}
\end{equation}
This constant $\sigma $ should be positive and independent of the incident
angle for a focus without aberration. Thus the following rule for the wave
vector refraction at the interface must be obeyed 
\begin{equation}
k_{rz}=\kappa -\sigma k_{z}.  \label{k-refrac}
\end{equation}
Here $\kappa $ is the integration constant which is also an intrinsic
property of the ENGRM. Since in the vacuum $k_{z}^2+k_{x}^2=k_{0}^2=\omega_{0}^2/c^2$, the
equi-frequency surface (EFS) of the medium is 
\begin{equation}
\sigma ^{-2}(\kappa -k_{rz})^{2}+k_{rx}^{2}=\omega _{0}^{2}/c^{2}.
\label{lens-EFS}
\end{equation}
Note that this elliptic dispersion exists only at the operating frequency $%
\omega _{0}$ \cite{veselago-comment}. In the neighborhood of this frequency $%
\omega \sim \omega _{0}$, $(\kappa
-k_{rz})^{2}/n_{z}^{2}+k_{rx}^{2}/n_{x}^{2}=\omega ^{2}/c^{2}$ 
with $\tilde{\sigma }\equiv
n_{z}/n_{x}\sim \sigma $. The deduction of Eq. (\ref{k-refrac}) from an
elliptic EFS requires \emph{negative group refraction} 
\begin{equation}
({\bf k}_{r}-\kappa {\hat{\bf{z}}})\cdot \nabla _{{\bf k}%
_{r}}\omega <0  \label{neg-group-ref}
\end{equation}
together with the causality condition $\partial \omega /\partial k_{rz}>0$
that the energy must flow away from the interface which is illustrated in
Fig. \ref{fig2}. The ray vector \cite{Landau} which represents the direction of group
velocity in a medium with elliptic EFS is 
${\bf s}=(-\tilde{\sigma }\sin \alpha {\hat{\bf{x}}}%
+\cos \alpha {\hat{\bf{z}}})/n_{x}\tilde{\sigma }$ while ${\bf k}%
_{r}-\kappa {\hat{\bf{z}}=}n_{x}(\sin \alpha {\hat{\bf{x}}}-%
\tilde{\sigma }\cos \alpha {\hat{\bf{z}}})$ with $\sin \alpha
=n_{x}^{-1}\sin \theta $ and $\theta$ the incident angle in the vacuum.

The Snell's law for wave vector loses its meaning in this anisotropic
medium. However there is a law for group refraction, which is 
$\tan \beta =-
\tilde{\sigma }\tan \alpha =-\tilde{\sigma }(n_{x}^{2}-\sin
^{2}\theta )^{-1/2}\sin \theta$.
At the operating frequency $\omega _{0}$
of the flat lens, $n_{x}=1$, $\tilde{\sigma }(\omega _{0})=\sigma $, the
group refraction law is simply 
\begin{equation}
\tan \beta =-\sigma \tan \theta .  \label{grp-refrac-law}
\end{equation}
\emph{Thus }$-\sigma $\emph{\ can be regarded as the effective refractive
index}. Note that this refraction law is very general and is valid for any $%
\sigma $ as defined by Eq. (\ref{sig-def}). The proof is the following, $%
\tan \beta =\frac{\partial \omega }{\partial k_{rx}}/\frac{\partial \omega }{%
\partial k_{rz}}=\frac{\mathrm{d}k_{rz}}{\mathrm{d}k_{z}}\frac{\partial
\omega }{\partial k_{x}}/\frac{\partial \omega }{\partial k_{z}}=-\sigma
\tan \theta $. Here we used $k_{rx}=k_{x}$.

The space inside the medium can be considered to be 
``optically stretched" in the $z$-direction by a factor of $%
\sigma ^{-1}$. The waves inside the medium are represented by 
$\psi ({\bf r}^{\prime })
=\exp(ik_{x}x-ik_{z}z^{\prime }+i\kappa ^{\prime }z^{\prime })$
with $z^{\prime }=\sigma z$ and $\kappa ^{\prime }=\kappa /\sigma $. Thus in
the case that $\kappa \neq 0$, the waves at this frequency will experience a
periodicity in the $z$-direction with length scale $2\pi /|\kappa |$ since
the waves are modulated by $\exp(i\kappa z)$. This periodicity is the
effective periodicity and may not be the actual physical periodicity such as
the lattice spacing of a PhC.

We now consider the imaging of an ENGRM slab with thickness $d$ as shown 
in Fig. \ref{fig1}(b). The first
surface is at the origin and the second at $z=d$. A point source\ is placed
at $z=-u$. We consider another point outside the ENGRM\ slab at $z=d+v$. If
the value of $v$ satisfies the following lens equation 
\begin{equation}
u+v=\sigma d,  \label{lens-1}
\end{equation}
the phase difference $\Phi _{z}=k_{z}(u+v)+k_{rz}d=\kappa d$ is stationary
and an image will be formed without aberration. Note that for an ordinary 
thin curved lens with focal length $f$, the lens equation is 
$u^{-1}+v^{-1}=f^{-1} $. With given value of $\sigma $ and slab thickness $d$%
, the distance between the object and the image outside the slab is fixed, $%
u+v+d=(1+\sigma )d$. Once the location of the object is fixed, the position
of the image is also fixed, no matter where the slab is placed!

We remark that a flat slab can focus with \emph{negative} as well as with 
\emph{\ positive }wave vector refraction. \emph{The necessary condition for
flat lens imaging is negative group refraction}. In general, there is not
much meaning of Snell's law for wave vectors for anisotropic media. Only the
refraction law for the direction of group velocity Eq. (\ref{grp-refrac-law}%
) is meaningful and does correspond to ray diagrams. \emph{The
Veselago-Pendry lens is a special case of our flat lens with} $\sigma =1$, $%
\kappa =0$ and is the only case that the ray diagram is applicable for the
wave vectors.

The lensmaker's formula Eq. (\ref{lens-0}) or (\ref{lens-1}) only provides a
necessary condition for imaging. \emph{An additional constraint is required
for the formation of a ``perfect" image,}
viz. the surface reflection coefficient $r$ should vanish, $r=0$, for far
field and should diverge, $r\rightarrow \infty $, for near field. These set
the conditions for a perfect flat lens without optical axis. The requirement
of no reflection $r=0$ is not essential for a slab to focus far field. \emph{%
The principal conditions for the flat lens to focus light are that Eq. (\ref%
{k-refrac}) and} $|r|\ll 1$ \emph{should be satisfied for all or at least a
large range of incident angles}. The presence of reflection will make the
image dim and may give rise to multiple images. The indefinite indices
medium \cite{Smith03} can have an elliptic dispersion relation, but in general
will not satisfy the flat lens imaging conditions.

The field inside the ENGRM flat lens can also be described by a partial
differential equation $[\sigma ^{-2}(i\kappa +\partial _{z})^{2}+\partial
_{x}^{2}+(\omega /c)^{2}]\Psi =0$. The medium can be regarded as a distorted
space with the metric 
$\mathrm{d}s^{2}=\mathrm{d}x^{2}+\sigma ^{2}\mathrm{d}z^{2}$. Thus there is an
analogy with gravitational lensing where refraction occurs due to the
warping of space caused by general relativistic effects, such as in the
vicinity of a massive object. It is worth noting that general relativistic
effects are observable due to optical refraction.

In the rest of the paper, we will focus on $S$-polarized electromagnetic
waves with the electric field in the $y$-direction. In general, the magnetic
permeability $\mu $ of the ENGRM should be a tensor, and its relationship to
the other material parameters $\sigma $ and $\kappa $ can be deduced. 
Here we assume that effective indices $\epsilon$ and $\mu$ can be used 
to describe the medium of interest. Due to
its symmetry, the ellipsoid can always be transformed to its principal axes 
\cite{BornWolf}. For the $S$-polarized waves, the $H$ field is in the $xz$%
-plane, only $\mu _{x}$ and $\mu _{z}$ will enter our discussion.
Furthermore, $\mu _{x}$ plays a more active role than $\mu _{z}$ since the
surface reflection coefficient is $r=(\mu _{x}k_{z}-k_{rz})/(\mu
_{x}k_{z}+k_{rz})$. Thus to achieve perfect imaging for far field, the
effective permeability of ENGRM should be 
\begin{equation}
\mu _{x}=\kappa /k_{0}\cos \theta -\sigma .  \label{mu-far}
\end{equation}
Only in the case $\kappa =0$ that $\mu _{x}=-\sigma $ is isotropic while for 
$\kappa \neq 0$, $\mu _{x}$ has to be anisotropic and diverges at
$k_{x}=k_{0}$. For $\kappa \neq 0$, any finite $\mu _{x}$ at $k_{x}=k_{0}$
results in reduced transmission. Notice that in PhCs including
metamaterials, the effective $\mu _{x}$ should be anisotropic in general.

\begin{figure}[tbp]
\center{\ \includegraphics [angle=0,width=3.2in]{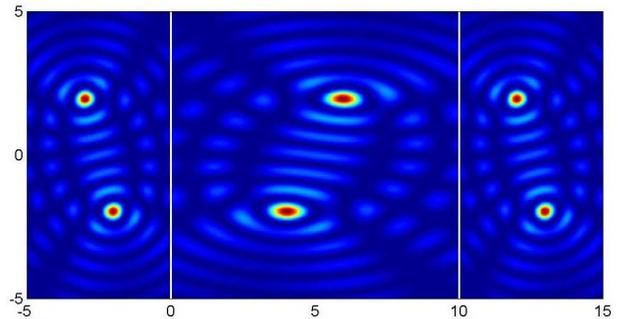}}
\caption{Field intensity of two sources ($J_0(k_0r)$) imaged by an ENGRM flat
lens with $\protect\sigma =0.5$, $\kappa=0$, and thickness $d=10$. 
The two white lines indicate the surfaces of the lens.
We set $\mu_x=-0.5$ for the ENGRM for perfect transmission.}
\label{fig3}
\end{figure}

To see how the propagating waves are focused, we consider a $S$-polarized 
point source $-(i/4)H^{(1)}_0(k_0|{\bf r}+u{\hat {\bf z}}|)$ which is centered
at $z=-u$ while the two surfaces of the ENGRM slab are at $z=0$ and $z=d$.
If $\mu _{x}$ takes the form given by Eq. (\ref{mu-far}), there will be no
reflection for propagating waves and the transmission coefficient through
the slab is $T=\exp(ik_{rz}d)$. The transmitted far field is 
\be
E_{\rm{far}}^{
\rm{t}}({\bf r})=-ie^{i\kappa d}\int_{0}^{\pi /2}
{\mathrm{d}\theta \over 2\pi}
e^{ik_{z}[z+u-(1+\sigma )d]}\cos (k_{x}x). 
\ee
Note that $k_z=k_0\cos\t$ and $k_x=k_0\sin\t$.
The image outside the slab is
located at $z=(1+\sigma )d-u$. One can see that except for a global phase 
$\exp(i\kappa d)$, the far field of the image is exactly that of the source.
The far field inside the slab is 
\be
E_{\rm{far}}^{\rm{in}}({\bf r})
=-ie^{i\kappa z}\int_{0}^{\pi /2}{\mathrm{d}\theta \over 2\pi} 
e^{-ik_{z}(\sigma z-u)}\cos (k_{x}x) 
\ee
for $0\leq z\leq d$. Another perfect image is formed inside
the slab at $z=u/\sigma $. The first surface acts as a mirror with an extra
phase $\exp(i\kappa z)$. The space inside the ENGRM slab is optically
stretched by a factor of $\sigma ^{-1}$. A flat lens imaging is shown in
Fig. \ref{fig3}.

\emph{Evanescent waves and sub-wavelength imaging }For evanescent waves $%
k_{x}>k_{0}$, the complex extrapolation of Eq. (\ref{lens-EFS}) gives 
\begin{equation}
k_{rz}=\kappa +i\sigma q
\end{equation}
with $q\equiv (k_{x}^{2}-k_{0}^{2})^{1/2}$. The transmission coefficient
through the flat lens is $T=(1-r^{2})\exp(ik_{rz}d)/[1-r^{2}\exp(2ik_{rz}d)]$
with $r=[(\mu _{x}-\sigma )q+i\kappa ]/[(\mu _{x}+\sigma )q-i\kappa ]$. In
order to amplify evanescent waves, a singularity must exist in $T$ to
compensate to certain extent the decay of evanescent waves in the vacuum 
$\exp[-q(u+v)]$. In the case of single-interface resonance \cite{Pendry,Luo03}, 
$r\rightarrow \infty $, $T\exp[-q(u+v)]=\exp(i\kappa d)$, all the evanescent
waves will be amplified only if 
\begin{equation}
\mu _{x}=\mu _{x}^{0}\equiv -\sigma +i\kappa /q.  \label{single-res}
\end{equation}
The images both inside and outside have the same sub-wavelength features of
the source. If $\mu _{x}$ couldn't take the above form, sub-wavelength
imaging is still possible. At the so-called overall resonance condition \cite%
{Luo03}, $1-r^{2}\exp(2ik_{rz}d)=0$, thus $r=\pm \exp(-ik_{rz}d)$, one gets 
\begin{equation}
\mu _{x}=\mu _{x}^{\pm }\equiv (-\sigma +i\kappa /q)\tanh ^{\mp 1}(\sigma
q-i\kappa )d/2.
\end{equation}
For $\kappa \neq 0$, $\mu _{x}^{0}$ and $\mu _{x}^{\pm }$ are all complex
with negative real parts. One notices that $\mu _{x}^{+}(q)$ is flatter than 
$\mu _{x}^{-}(q)$ as functions of $q$. In terms of effective indices, the
existence of surface modes $\omega _{0}(k_{x})$ requires that $\Re \mu _{x}$
be negative for $S$-polarization. To amplify evanescent waves, the $\omega
_{0}(k_{x})$ curve must be very flat \cite{Luo03}. This is equivalent to say
that the curve $\mu _{x}^{+}(q)$ should be flat so that the contribution of
evanescent waves are constructive for certain window of $q$. The closer $\mu
_{x}$ is to $\mu _{x}^{+}(q)$, the more sub-wavelength features the images
will have. Note that for large $\kappa d$, the evanescent waves are located
in the vicinity of the interfaces for constant $\mu _{x}$.

\emph{Realization in real materials} We note that no known natural material
is found to have negative dispersion. Negative group refraction can be
achieved in periodic or quasi-periodic media with nonzero $\kappa $ \cite%
{NotePhC}. \emph{For most media, Eq. (\ref{mu-far}) and (\ref{single-res})
are unlikely to be fulfilled, thus a perfect flat lens is unattainable.}
Realization using real materials is discussed next.

\begin{figure}[tbp]
\center{\ 
\includegraphics [angle=0,height=2.6in]{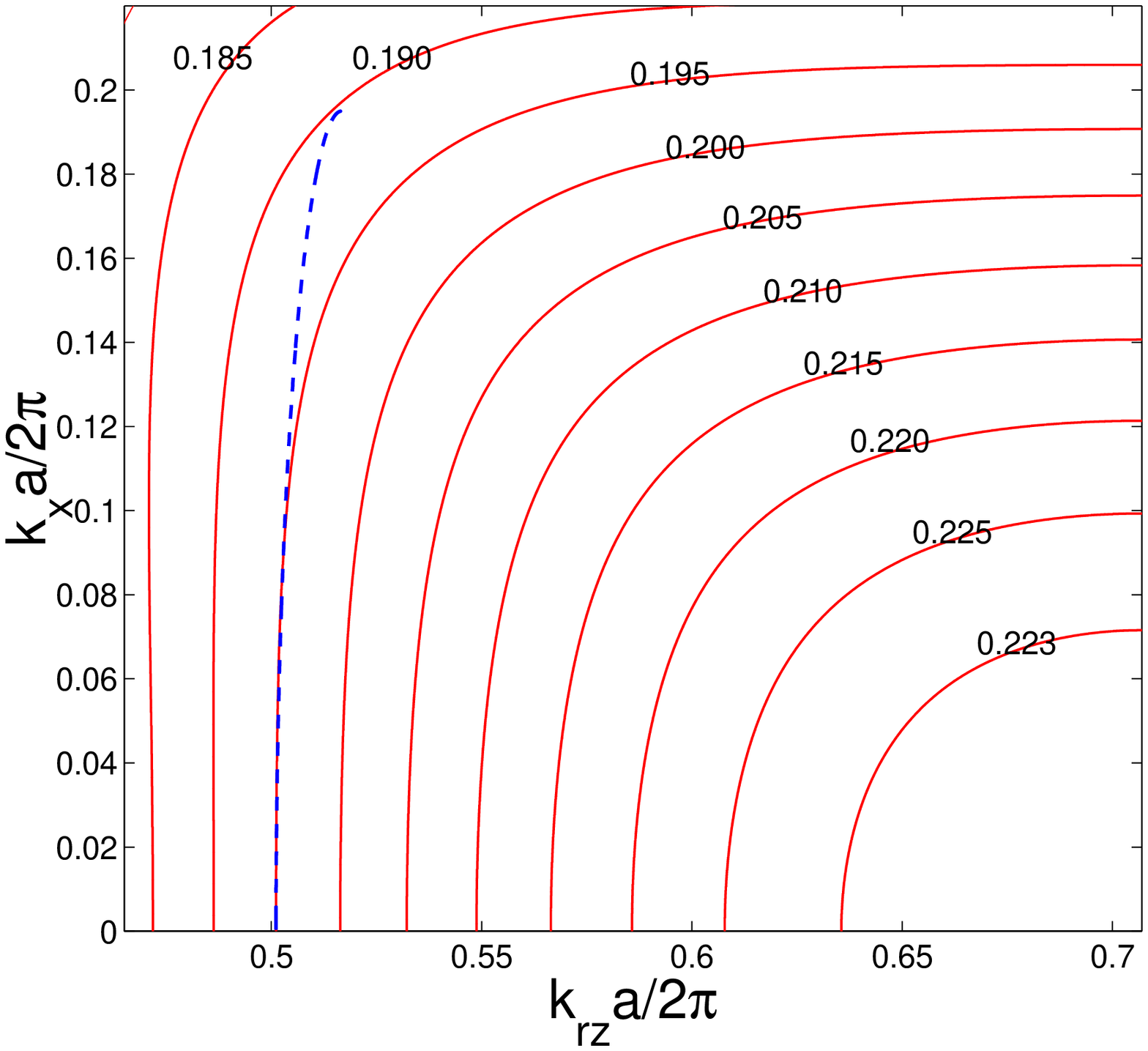} 
\includegraphics [angle=0,height=2.3in]{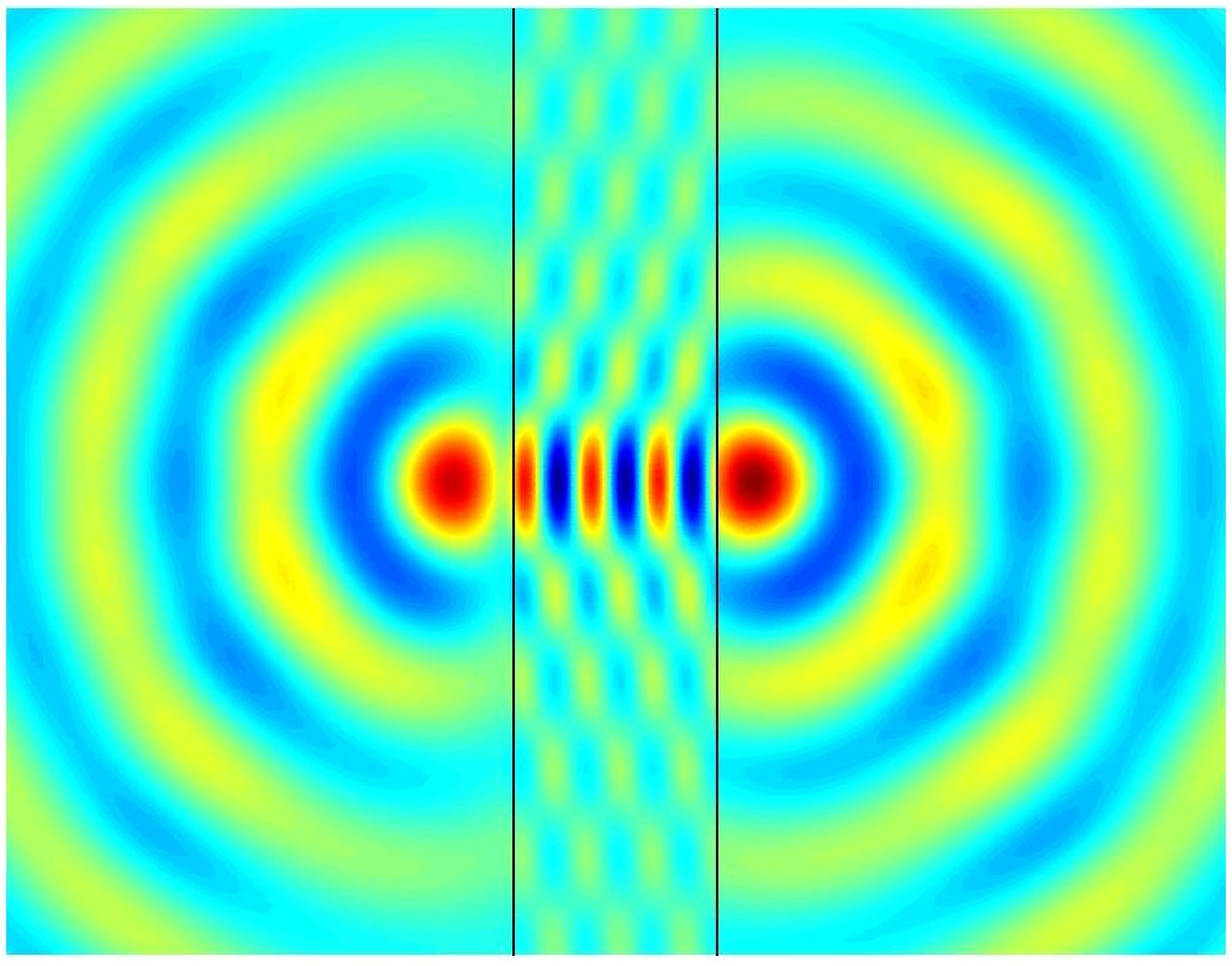}}
\caption{(a)\ EFS of the TE modes of a square lattice PhC calculated using
plane wave expansion. (b) Far-field $H_{z}$ of a point source across an
ENGRM slab with dispersion relation $k_{rz}=\protect\kappa -\protect\sigma %
k_{z}$ with $\protect\kappa =3.2465$ and $\protect\sigma =0.08$ which
approximates the actual EFS of the PhC (dashed line in the upper panel). The
permittivity $\protect\epsilon _{x}=2.2$ is used. Note the modulated field
inside the homogeneous ENGRM. Due to impedance mismatch, the images are not
``perfect" unlike Fig. \ref{fig3}.}
\label{fig4}
\end{figure}

Ordinary material may not have the EFS described by Eq. (\ref{lens-EFS}) for
all $k_{x}$. For an uniaxial crystal, since the group refraction is
positive, thus $k_{rz}=\kappa -\sigma k_{z}$ with $\kappa =0$ and $\sigma<0$. 
There will be no focus though negative refraction for certain incident angles 
can be easily achieved \cite{Zhang03}. Instead a virtual image will be formed
satisfying the same lens equation. However, a PhC could have \emph{both} an
elliptic EFS and negative group refraction for certain windows of $k_{x}$ at
certain frequency. Here we use some general features of PhCs to explain the
imaging mechanism. To this end, we consider the first band of a square
lattice PhC of unit lattice spacing. For small $k_{x}$ along the $\Gamma M$
direction, the dispersion can be approximated elliptically as $k_{rz}\simeq
\kappa -\sigma _{0}k_{z}$ with appropriate constants $\kappa $ and $\sigma
_{0}$ (see the dashed line in Fig. \ref{fig4}). 
At $\omega _{0}=\omega _{u}$, the upper limit of
all-angle negative refraction (AANR) in the first band \cite{Luo02}, $\kappa =\sqrt{2}\pi
-k_{0}(1-\sigma _{0})$. Since for AANR with $\omega_{0}$ away from $\omega_{u}$, 
$\sigma_{0}$ will be smaller, we only concentrate here on the discussion for 
$\omega _{0}=\omega _{u}$. To enhance the transmission, the slab thickness can
be selected to satisfy the Fabry-Perot resonance condition $(\kappa -\sigma
_{0}\omega /c)d=0\ (\mathrm{mod}\ 2\pi )$. The slab is on resonance only for 
$k_{x}\rightarrow 0$. For large $k_{x}$, $k_{rz}\simeq \sqrt{2}\pi -\sigma
_{0}^{-1/2}k_{z}$ due to the $C_{4v}$ symmetry of the EFS. A desired large 
transmission would require $\mu _{x}$ to diverge.
Thus the image location is bounded $\sigma _{0}\leq (u+v)/d\leq \sigma
_{0}^{-1/2}$ and the Fourier components of the object with $k_{x}\rightarrow
k_{0}$ will be partially lost. This leads to a so-called \emph{%
self-collimation effect} for small\emph{\ }$\sigma _{0}$. Far field images
both inside and outside the slab with reasonably good quality will be
formed. However, the image inside the PhC is stretched and modulated due to
the partial transmission besides the Bloch wave modulation. Even with the removal
of Bloch wave modulation, the image inside a PhC will be visible only for large
thickness. For evanescent waves at $\omega =\omega _{u}$ with $%
k_{x}\rightarrow k_{0}$, one has $\Re \mu _{x}^{0}=-\sigma _{0}^{-1/2}$ and $%
\Im \mu _{x}^{0}\rightarrow \infty $ while for $k_{x}\gg k_{0}$, $\mu
_{x}^{\pm }(k_{x})\simeq \mu _{x}^{0}(k_{x})\rightarrow -\sigma _{0}^{-1}$. To
have substantial sub-wavelength feature of the image, the real part of $\mu
_{x}$ must take large negative value for small $\sigma _{0}$. An example is
shown in Fig. \ref{fig4}, which is very similar to Fig. 5 in Ref. \cite{Luo02}. The
image is confined in the vicinity of the second surface of the PhC slab due
to a small value of $\sigma _{0}$.

In this paper, we have discussed the group refraction of an anisotropic
medium characterized by two materials parameters $\sigma $ and $\kappa $.
When $\sigma $ is a positive constant, the anisotropic medium is an ENGRM and a
flat slab of such medium can be used as a focal lens without optical axis
and leading to images free of aberration. Our theory of flat lens imaging is
a generalization of the Vesalago-Pendry perfect lens and beyond Silin's
formula. The theory is valid for both real and virtual images. This flat lens
theory leads to a clear understanding of negative group refraction and flat
lens imaging in electromagnetism \cite{Parimi03,Luo02}, acoustics \cite%
{ZhangLiu}, and electron waves. Numerical simulations of homogeneous ENGRMs
and of PhCs were carried out supporting our theory. 

The lack of optical axis for the flat lens has very broad applications and
confers important advantages in optics. Clearly there are no aperture
restrictions. There are two characteristics that might be viewed as
limitations. The lensmaker's formula clearly dictates upper bounds to $%
u,v<\sigma d$ for real image formation. The other limitation is that
magnification is always unity. For any medium the lens properties $\sigma $, 
$\kappa $, and also the working frequency $\omega _{0}$ can be obtained by
inspection of the EFS. To have high quality image, $\sigma $ should be constant
and close to unity. The present theory can be used to design tailor-made
flat lenses. Extension of our theory to three-dimensions is straightforward.

Work supported by the National Science Foundation and the Air Force Research
Laboratories, Hanscom, MA.



\begin{thebibliography}{99}
\bibitem{veselago-comment} Note that the Veselago-Pendry lens also operates
at a single frequency for which $n=\varepsilon =\mu =-1$.
\bibitem{NotePhC} For a PhC with negative group refraction, $\kappa $ is
always nonzero. For example for the second band, $\kappa \sim 2\pi /a$
instead of $\kappa \sim 0$.
\end{thebibliography}
\end{document}